\documentclass[11pt]{article}
\usepackage{graphicx}
\usepackage{subfigure}
\usepackage{epsfig}
\usepackage{multirow}

\newcommand{\BABARPubYear}    {06}

\newcommand{\BABARConfNumber} {014}
\newcommand{\SLACPubNumber} {11974}
\newcommand{\LANLNumber} {0607040}

\input babarsym

\long\def\inst#1{\par\nobreak\kern 4pt\nobreak
    {\it #1}\par\vskip 10pt plus 3pt minus 3pt}

\begin{document}
{\pagestyle{empty}

\begin{flushright}
%BAD 1563 V6\\
\babar-CONF-\BABARPubYear/\BABARConfNumber \\
%\babar-PUB-\BABARPubYear/\BABARPubNumber \\
SLAC-PUB-\SLACPubNumber \\
hep-ex/\LANLNumber \\
%July 2006 \\
\end{flushright}

\par\vskip 5cm

% Title of the paper
\begin{center}
\Large \bf Search for the Baryon and Lepton Number Violating Decays {\boldmath$\tau \to \Lambda h$}
\end{center}
\bigskip

\begin{center}
\large The \babar\ Collaboration\\
\mbox{ }\\
\today
\end{center}
\bigskip \bigskip

% Abstract
\begin{center}
\large \bf Abstract
\end{center}
We have searched for the violation of baryon number $B$ and lepton
number $L$ in the $(B\!-\!L)$-conserving modes $\tau^- \to \bar{\Lambda^0}
\pi^-$ and $\tau^- \to \bar{\Lambda^0} K^-$ as well as the
$(B\!-\!L)$-violating modes $\tau^- \to \Lambda^0 \pi^-$ and $\tau^- \to
\Lambda^0 K^-$ using 237 fb$^{-1}$ of data collected with the \babar\
detector at the PEP-II asymmetric-energy $e^+e^-$ storage ring.  We do
not observe any signal and determine preliminary upper limits on the
branching fractions ${\cal B}(\tau^- \to \bar{\Lambda^0} \pi^-) < 5.9
\times 10^{-8}$, ${\cal B}(\tau^- \to \Lambda^0 \pi^-) < 5.8 \times
10^{-8}$, ${\cal B}(\tau^- \to \bar{\Lambda^0} K^-) < 7.2 \times
10^{-8}$, and ${\cal B}(\tau^- \to \Lambda^0 K^-) < 15 \times 10^{-8}$
at $90\%$ confidence level.  \vfill
\begin{center}

Submitted to the 33$^{\rm rd}$ International Conference on High-Energy Physics, ICHEP 06,\\
26 July---2 August 2006, Moscow, Russia.

\end{center}

\vspace{1.0cm}
\begin{center}
{\em Stanford Linear Accelerator Center, Stanford University, 
Stanford, CA 94309} \\ \vspace{0.1cm}\hrule\vspace{0.1cm}
Work supported in part by Department of Energy contract DE-AC03-76SF00515.
\end{center}

\newpage
} % end of pagestyle{empty}

% Input author list file
%
%author list removed temporarily to save trees 7/9/04 RNC
%
\begin{center}
\small

The \babar\ Collaboration,
\bigskip

%% author list as of 01-Jul-2006 (596 authors)
%
{B.~Aubert,}
{R.~Barate,}
{M.~Bona,}
{D.~Boutigny,}
{F.~Couderc,}
{Y.~Karyotakis,}
{J.~P.~Lees,}
{V.~Poireau,}
{V.~Tisserand,}
{A.~Zghiche}
\inst{Laboratoire de Physique des Particules, IN2P3/CNRS et Universit\'e de Savoie,
 F-74941 Annecy-Le-Vieux, France }
% ...

\end{center}\newpage

% The body of the paper starts here
\section{INTRODUCTION}
\label{sec:Introduction}

One of the important unresolved questions of our time is the presence
of a large baryon asymmetry in today's universe. According to
A. Sakharov \cite{sakharov} three conditions must be satisfied in
order for a baryon asymmetry to arise from an initial state with zero
baryon number: baryon number violation, C and CP symmetry violation,
and a departure from thermal equilibrium. No baryon number violating
processes have yet been observed \cite{PDG}. Though we know that the
baryon number was violated in the early universe we do not know how it
came about. Conservation of angular momentum requires that the spin
$1/2$ of a nucleon that is decaying to a lepton be transferred to the
lepton: $\Delta B= \pm \Delta L$. Therefore there are two types of
baryon instabilities $|\Delta(B\!-\!L)|=0,2$. In the Standard Model (SM),
and in most of its extensions, it is required that
$\Delta(B\!-\!L)=0$. The second possibility of $|\Delta(B\!-\!L)|=2$ allows
transitions with $\Delta B= - \Delta L$, or $|\Delta B|=2$ and
$|\Delta L|=0$, or $|\Delta L|=2$ and $|\Delta B|=0$. It follows that
the conservation or violation of $(B\!-\!L)$ determines the mechanism of
baryon instability.

It has been shown that, in baryogenesis, nonperturbative Standard
Model effects at the electroweak energy scale will erase any baryon
excess generated by $(B\!-\!L)$-conserving processes at the earliest
moments of the universe ($T>>1$~TeV) \cite{kuzmin}. In addition,
generating a baryon excess through electroweak effects alone does not
seem to be adequate to account for the observed baryon asymmetry
\cite{rubakov}. A component with $\Delta(B\!-\!L)=2$ might be necessary to
explain baryogenesis.

Most existing searches for $(B\!-\!L)$ violation have been restricted to
experiments with nucleons~\cite{PDG}. In this analysis we search for
the decays $\tau \to \Lambda \pi$ and $\tau \to \Lambda K$ , in the
$(B\!-\!L)$-conserving modes $\tau^- \to \bar{\Lambda^0} \pi^-(K^-)$ as
well as the $(B\!-\!L)$-violating modes $\tau^- \to \Lambda^0
\pi^-(K^-)$. Charge conjugate modes are always included if not
mentioned otherwise. A similar analysis of the modes $\tau \to \Lambda
\pi$ published recently by the Belle
Collaboration~\cite{belleLambdaPi} finds the upper limits ${\cal
B}(\tau^- \to \bar{\Lambda^0}\pi^-) < 14 \times 10^{-8}$ and ${\cal
B}(\tau^- \to \Lambda^0 \pi^-) < 7.2 \times 10^{-8}$ at $90\%$
confidence level (C.L.).

Experimental limits on the proton lifetime imply that the expected
branching fraction for $\tau \to (\bar{p} + \mbox{anything})$ is not
observable in the Standard Model: ${\cal B}(\tau \to \bar{p} + X) <
10^{-40}$ \cite{marciano}.  The $\Lambda^0$ baryon couples weakly to
the proton. We would then expect similar but approximately $10^{8}$
times weaker \cite{marciano} constraints from the proton lifetime for
$\tau \to \Lambda \pi(K)$.  A recent theoretical paper \cite{hou}
studied dimension-6 operators and concludes that baryon number
violation in decays involving higher generations, assuming proton
stability, will not be observable. However such a model may not be
adequate to describe the apparent baryon asymmetry in the first place.
Models with dimension-9 operators and yet unknown mechanisms that
generate baryon number violation or enhance the coupling to higher
generations may be able to accomplish this \cite{kamyshkov}.

With the advent of the $B$ factories, that also produce large
quantities of $\tau$ leptons, we are now able to experimentally study
such decays with greatly improved precision.

\section{THE \babar\ DETECTOR AND DATASET}
\label{sec:babar}

This measurement was performed using data collected by the
\babar~detector at the PEP-II storage ring. Charged particles are
detected and their momenta measured by a combination of a silicon
vertex tracker (SVT), consisting of 5 layers of double-sided
detectors, and a 40-layer central drift chamber (DCH), both operating
in a 1.5-T axial magnetic field. Charged particle identification is
provided by the energy loss in the tracking devices and by the
measured Cherenkov angle from an internally reflecting ring-imaging
Cherenkov detector (DIRC) covering the central region. Photons and
electrons are detected by a CsI(Tl) electromagnetic calorimeter (EMC).
The EMC is surrounded by an instrumented flux return (IFR).  Electrons
are identified using measurements from the DCH, EMC, and DIRC. The
average identification efficiency is approximately $97\%$, whereas the
pion (kaon) misidentification rate is less than $2\%$ ($1\%$). Kaons
are identified using the SVT, DCH, and DIRC. The average
identification efficiency for the tight kaon selection is
approximately $80\%$, whereas the pion misidentification rate is less
than $1\%$. The average identification efficiency for the loose kaon
selection is approximately $90\%$, whereas the pion misidentification
rate is less than $4\%$. Protons are identified with a likelihood
based algorithm using measurements from all described detector
components. The proton identification efficiency ranges from
approximately $90\%$ to $96\%$ depending on polar angle and momentum,
whereas the average pion (kaon) misidentification rate is $5\%$
($12\%$). Details of the detector are described elsewhere \cite{nim}.

The data sample used corresponds to an integrated luminosity of 237
fb${}^{-1}$ collected from $e^+e^-$ collisions at, or $40$~MeV below,
the $\Upsilon(4S)$ resonance. Production and decay of the tau leptons
are simulated with the kk2f \cite{kk2f,kk2f2} and tauola
\cite{tauola,tauola2} Monte Carlo (MC) event generators, according to
two-body phase space, and taking spin correlations into account for
the signal mode. $B$ meson decays are simulated with the EvtGen
generator \cite{evtgen}, and $q\bar{q}$ events, where $q=u,d,s,$ or
$c$ quark, with the JETSET \cite{jetset} generator. The detector is
fully modelled using the GEANT4 simulation package \cite{geant4}.

\section{ANALYSIS METHOD}

We reconstruct candidate events $e^+e^- \to \tau^+\tau^-$ with one
$\tau$ decaying to $\Lambda \pi(K)$ and $\Lambda \to p \pi$. The other
tau in each event is required to be a one-prong decay. Decays that
conserve $(B\!-\!L)$ are recognized by opposite sign charge of the pion or
kaon from the $\tau$ decay and the pion from the $\Lambda^0$ decay. In
decays where $(B\!-\!L)$ is violated the two charges have the same sign.

Each event must have exactly four well reconstructed tracks in the
fiducial volume of the DCH with a total charge of zero. We divide the
events into two hemispheres defined by the thrust axis of the
event. The thrust axis is calculated using tracks in the drift chamber
and calorimeter energy depositions without an associated track. We
require that the three signal tracks are contained in one hemisphere
and that there is exactly one remaining track in the other hemisphere,
which we will refer to as the tagging hemisphere.

One of the signal tracks must be identified as a proton and, when
combined with an oppositely charged signal track, must give a $p
\pi^-$ invariant mass within $5$~MeV$/c^2$ of the nominal $\Lambda^0$ mass
\cite{PDG}. The set of signal tracks are subjected to a topological
fit to the decay tree $\tau \to \Lambda \pi (K)$, which must converge
and return a $\chi^2$ probability greater than $2.5\%$.

We require that the center-of-mass (CM) momentum of the $\Lambda^0$ is
greater than the lower kinematic limit of $1.8$~GeV/$c$ for $\tau^-
\to \Lambda^0 \pi^-$ decays. A requirement on the $\Lambda^0$ flight
distance $L_{\Lambda^0}>1$~cm and the signed flight length
significance $L_{\Lambda^0}/\sigma_{\Lambda^0}>0$ removes
$\tau^+\tau^-$ ($88\%$) and $q\bar{q}$ ($22\%$) events that do not
contain true $\Lambda$ particles. The remaining backgrounds are mostly
from $q\bar{q}$ events and to a lesser degree $\tau^+\tau^-$ events
that contain $K^0_{\rm s}$ decays and photon conversions $\gamma \to
e^+e^-$. None of approximately $800$ million MC $B\bar{B}$ events
survive the selection criteria.

\begin{figure}[!tb]
\unitlength1cm
  \begin{center}
    \subfigure[]{\epsfig{file=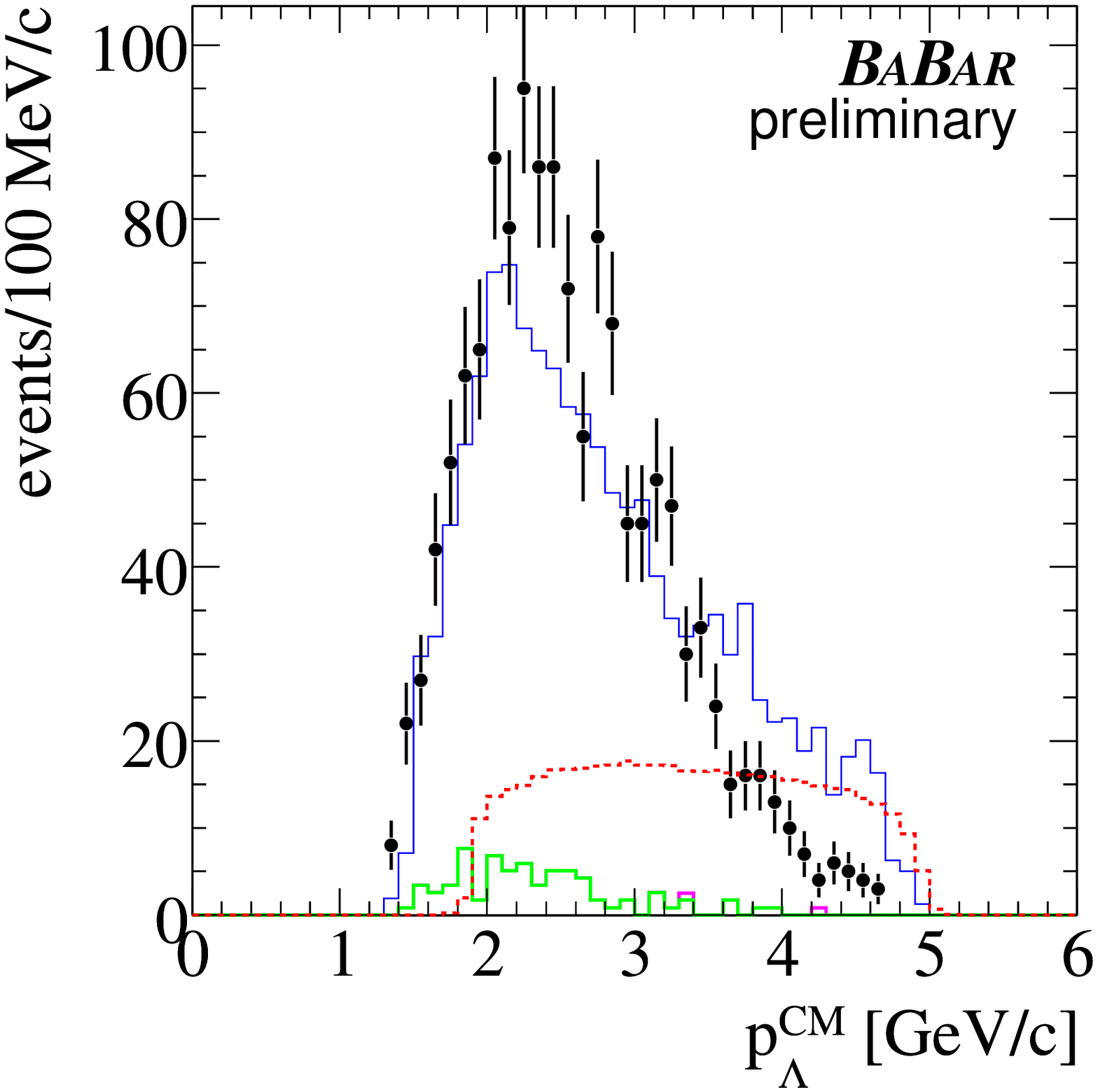,width=6cm}}
    \subfigure[]{\epsfig{file=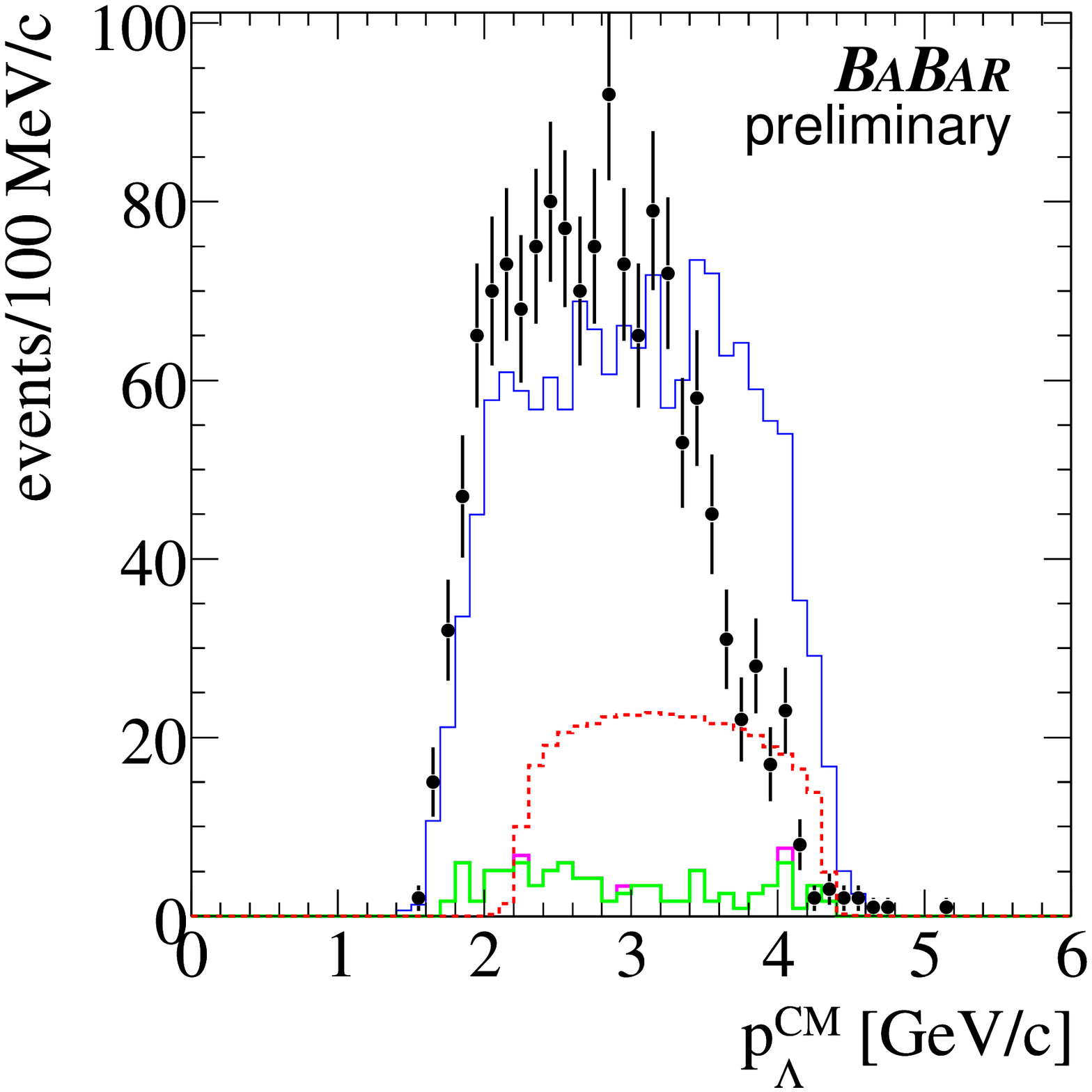,width=6cm}}
    \subfigure[]{\epsfig{file=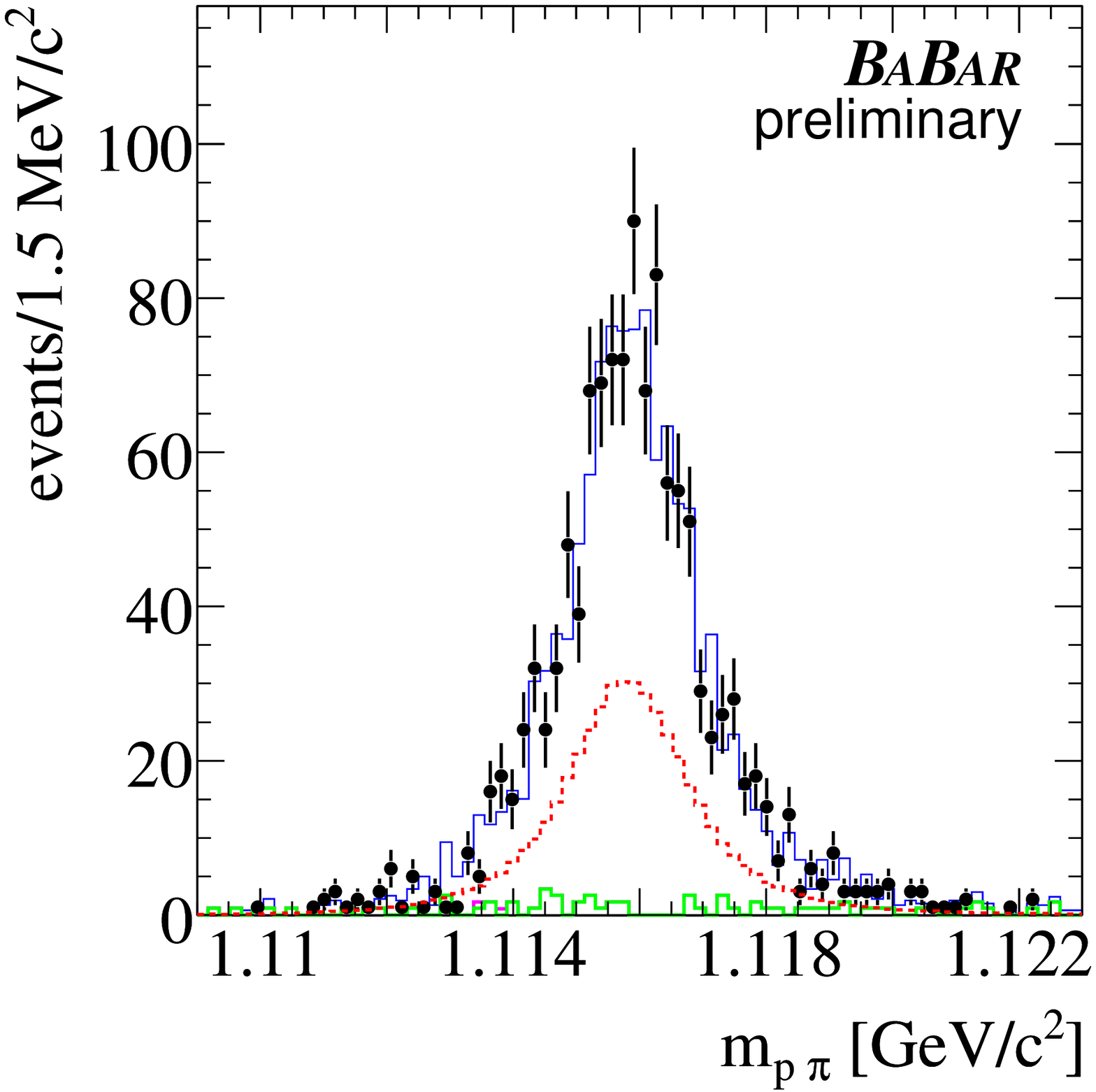,width=6cm}}
    \subfigure[]{\epsfig{file=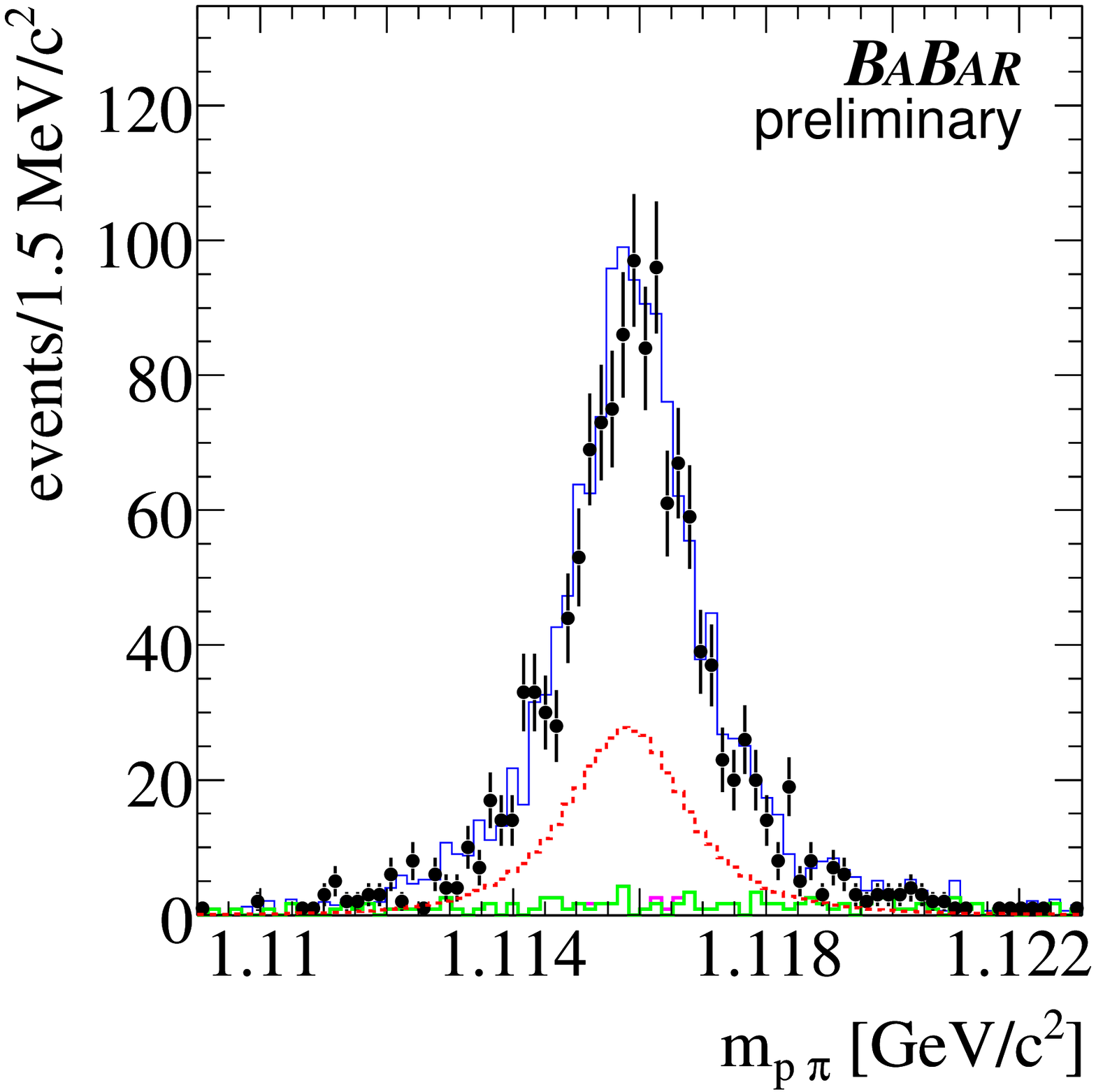,width=6cm}}
    \caption{
    The $\Lambda^0$ candidate momentum in the $e^+e^-$ rest frame for (a) $\tau \to \Lambda \pi$ and (b)  $\tau \to \Lambda K$ and
    the $\Lambda^0$ invariant mass spectrum for (c) $\tau \to \Lambda \pi$ and (d)  $\tau \to \Lambda K$.
    The solid stacked histograms are from top to bottom: $uds$ backgrounds (blue), and $\tau^+\tau^-$ (green). 
    The $c\bar{c}$ component is too small to be seen in these figures.
    The signal Monte Carlo distributions are shown with the dashed red histogram. The points correspond to events in the data sidebands.}
    \label{figCompare}
  \end{center}
\end{figure}

Figure \ref{figCompare} shows a comparison of the MC simulation with
our data. Note that the $\Lambda^0$ momentum spectrum shown in Figure
\ref{figCompare} (a,b) is not very well described by our MC
simulation. This is most likely due to imperfections of the $q\bar{q}$
MC event generator. For this reason the final background will be
determined from the data. All other variables that were studied show
better agreement between data and MC.

We require that the pion track from the $\Lambda^0$ decay as well as
the tagging track from the other $\tau$ lepton do not pass tight kaon
identification requirements. In the mode $\tau \to \Lambda \pi$ we
require that the $\pi$ is not identified as a kaon. In the mode $\tau
\to \Lambda K$ we require that the kaon track be identified with loose
kaon identification requirements.  To suppress candidates that include
tracks from photon conversions, we require that the pion or kaon from
the $\tau$ decay and the pion from the $\Lambda^0$ decay must not be
identified as an electron. The pion or kaon from the $\tau$ decay must
not be identified as a proton.

We study events in the two dimensional plane $m_{\Lambda \pi(K)}$
versus $\Delta E_{\Lambda \pi(K)}$, where $m_{\Lambda \pi(K)}$ is the
invariant mass of the $\Lambda$ and the pion (or kaon) candidate, and
$\Delta E_{\Lambda \pi(K)}= E_{\Lambda \pi(K)} - \sqrt{s}/2$ is the
reconstructed energy $E_{\Lambda \pi(K)}$ of the signal tracks minus
the expected $\tau$ energy, which is half the known $e^+e^-$
center-of-mass energy $\sqrt{s}$. A rectangular region that includes
the signal region was blinded during the development of this
analysis. Signal candidates are counted in an elliptical signal region
with a half width of $10$~MeV in $m_{\Lambda \pi(K)}$ and $90$~MeV in
$\Delta E_{\Lambda \pi(K)}$ centered around the nominal $\tau$ mass
\cite{PDG} and $\Delta E_{\Lambda \pi(K)}=0$. In the case of $\tau \to
\Lambda K$ the width in $m_{\Lambda \pi(K)}$ is reduced to $7$~MeV
because of the better resolution in this mode. The elliptical signal
region is slightly tilted to reflect the small correlation between the
two variables. The tilt is $\approx 3^{\circ}$, which can also be
expressed as a correlation coefficient between the two variables:
$\rho= 0.42$ for $\tau \to \Lambda \pi$ and $\rho=0.56$ for $\tau \to
\Lambda K$. The definition of the signal region as well as the other
selection requirements applied in this analysis have been optimized
using MC simulation, to obtain the lowest average upper limit for the
signal modes under the assumption that no signal will be observed.

\begin{figure}[!htb]
  \unitlength1cm
  \begin{center}
    \subfigure[$\tau \to \Lambda \pi$]{\epsfig{file=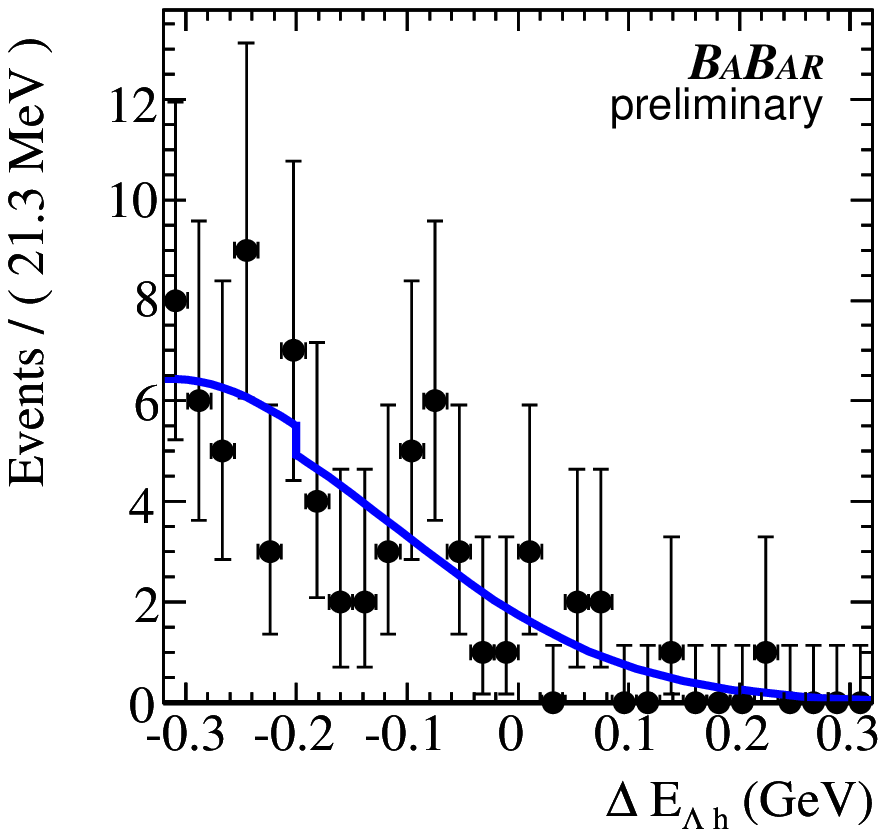,width=6cm}}
    \subfigure[$\tau \to \Lambda K$]{\epsfig{file=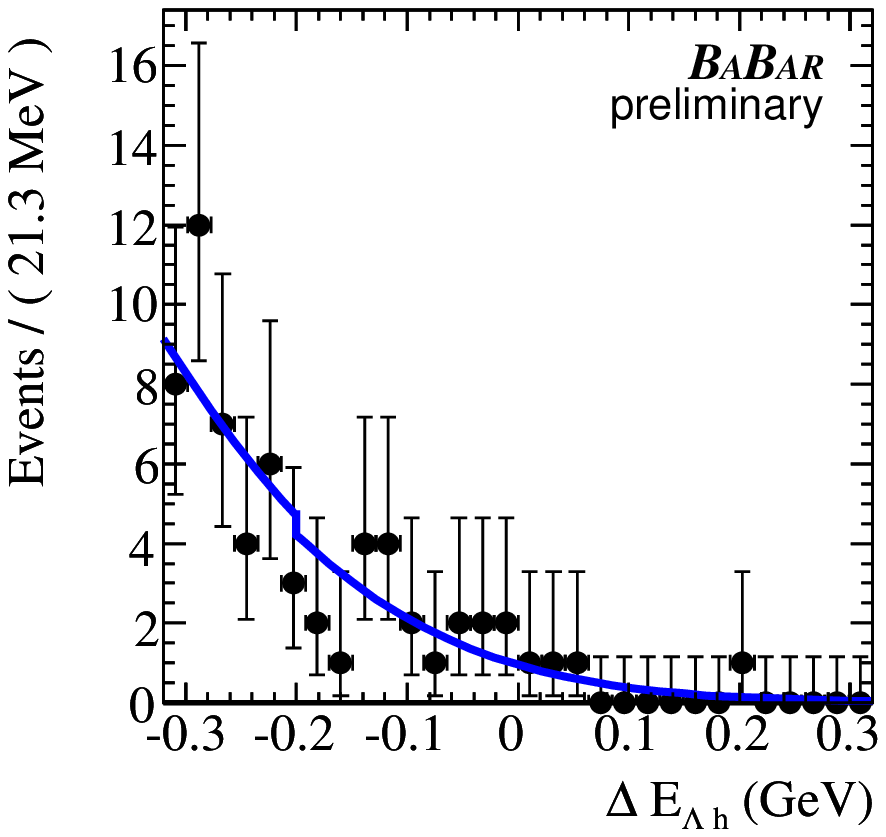,width=6cm}}
    \subfigure[$\tau \to \Lambda \pi$]{\epsfig{file=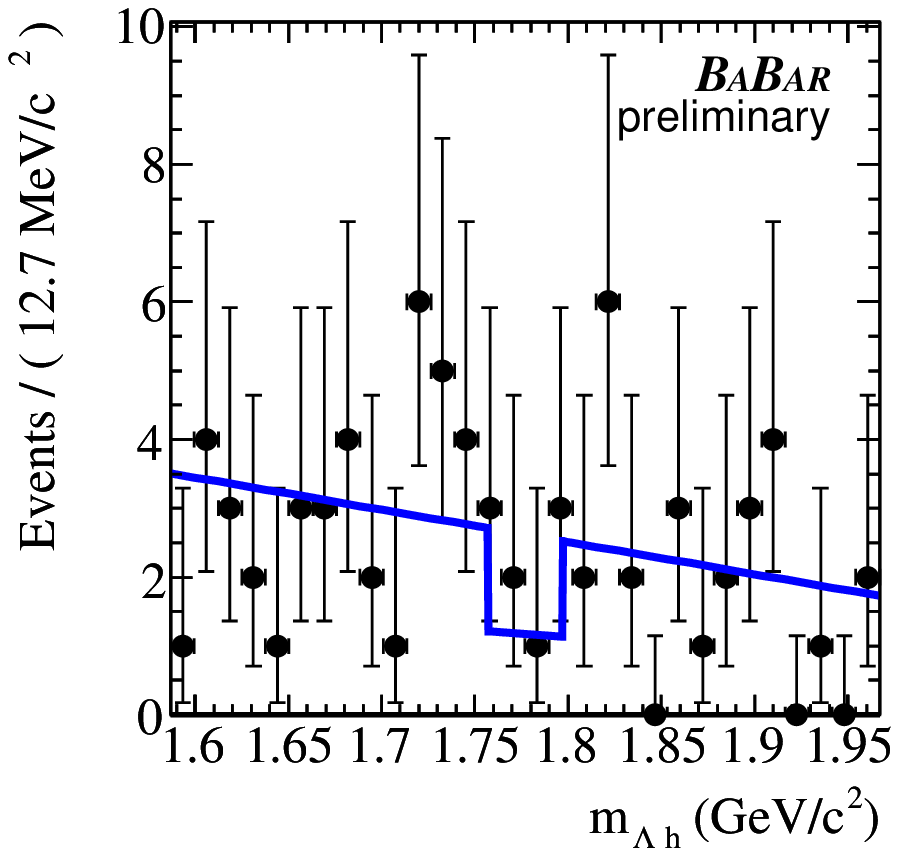,width=6cm}}
    \subfigure[$\tau \to \Lambda K$]{\epsfig{file=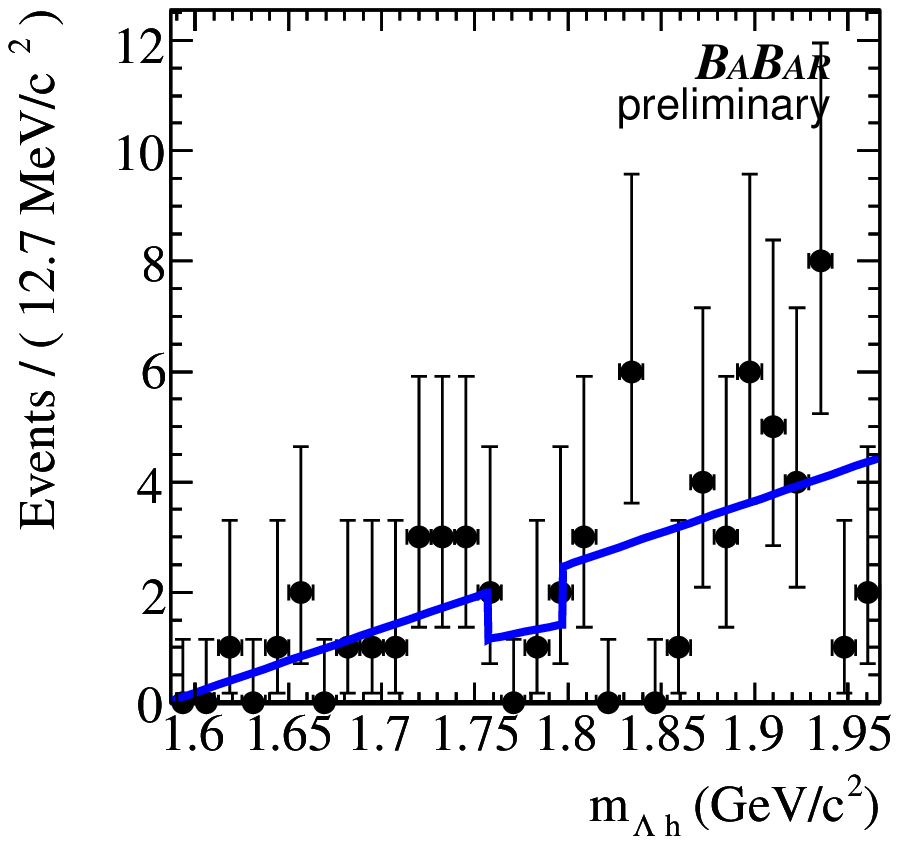,width=6cm}}
    \caption{
     Projections of the background parametrization as derived from a 2D unbinned maximum likelihood fit. The
     top row shows the $\Delta E$ projection, and the bottom row the $m_{\Lambda h}$ projection.
     The $\tau \to \Lambda \pi$ mode is shown in the left column and the $\tau \to \Lambda K$ mode
     in the right column. The fitted probability density function (PDF) is indicated by a line. The PDF was required
     to be zero in the blinded region, which causes the apparent drop around the signal regions in these projections.
     The points with error bars correspond to $\tau \to \Lambda \pi(K)$ candidates in data, outside the blinded region.}
    \label{figdEData}
  \end{center}
\end{figure}

We estimate the number of background events in the signal region with
a 2D unbinned maximum likelihood fit of the $m_{\Lambda \pi(K)}$ and
$\Delta E_{\Lambda \pi(K)}$ distributions outside the blinded region.
We try a number of functional forms that describe both the data and MC
distributions. The default fit uses a simple parametrization that
describes the data well and results in a background estimate that is
in the center of the possible range of values. A first-order
polynomial is fitted to the $m_{\Lambda \pi(K)}$ distribution and a
Gaussian function to the $\Delta E_{\Lambda \pi(K)}$ distribution. The
blinded region is excluded from the fit and the probability density
function is set to zero within the blinded region. The
parametrizations obtained are shown in Figure \ref{figdEData}. The
elliptical signal regions and the blinded region are also indicated in
Figure \ref{figdEvsM2d}. Due to the uncertainties of the background
parametrization and the possibility of correlations among the fit
variables, we take a conservative $100\%$ error on the number of
estimated background events in the signal region.

\section{SELECTION EFFICIENCY}
\label{sec:Systematics}

The signal efficiencies have been obtained from Monte Carlo
simulations.  Systematic uncertainties have been studied using
independent control samples of real data; a summary is presented in
Table \ref{tblSystematics}.  The largest contributions are from
uncertainties related to the tracking efficiency, and $\Lambda$
reconstruction. The latter has been estimated by comparing lifetime
distributions of long lived particles in data and Monte Carlo.  The
uncertainty on the branching fraction ${\cal B}(\Lambda^0 \to p
\pi^-)$ has been taken from the Review of Particle Physics
\cite{PDG}. Contributions to the systematic uncertainty are added in
quadrature to give a total systematic uncertainty of $6.9\%$ in the
mode $\tau \to \Lambda \pi$ and $7.0\%$ for $\tau \to \Lambda K$.

\begin{table}[!htb]
\begin{center}
\caption{Summary of systematic uncertainties on the signal efficiency, and
the luminosity and cross section.}
\vspace{0.2cm}
\begin{tabular}{lc}
\hline \hline
source                   & uncertainty $(\%)$ \\
\hline
$\Lambda$ reconstruction      & $5.0$ \\
tracking efficiency           & $4.0$ \\
proton identification         & $1.0$ \\
kaon identification ($\tau \to \Lambda K$ only) & $1.0$ \\
${\cal B}(\Lambda \to p \pi)$ & $0.8$ \\
\hline
luminosity and cross section  & $2.3$ \\
\hline
total $\tau \to \Lambda \pi$   & $6.9$ \\
total $\tau \to \Lambda K$     & $7.0$ \\
\hline \hline
\end{tabular}
\label{tblSystematics}
\end{center}
\end{table}

\section{RESULTS}
\label{sec:Physics}

\begin{figure}[!tb]
  \unitlength1cm
  \begin{center}
    \subfigure[$\tau^- \to \bar{\Lambda^0} \pi^-$]{\epsfig{file=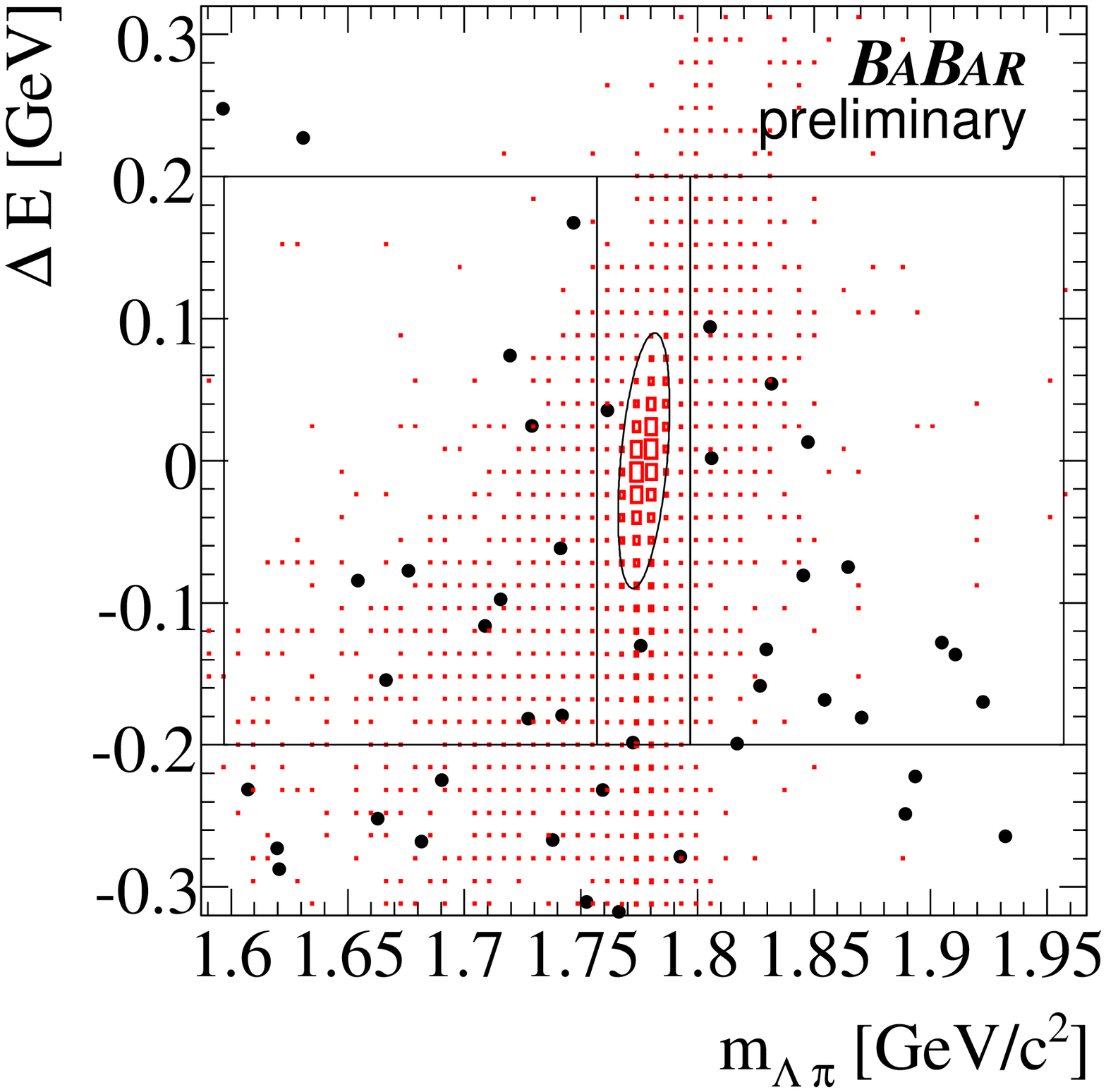,width=6cm}}
    \subfigure[$\tau^- \to \Lambda^0 \pi^-$]{\epsfig{file=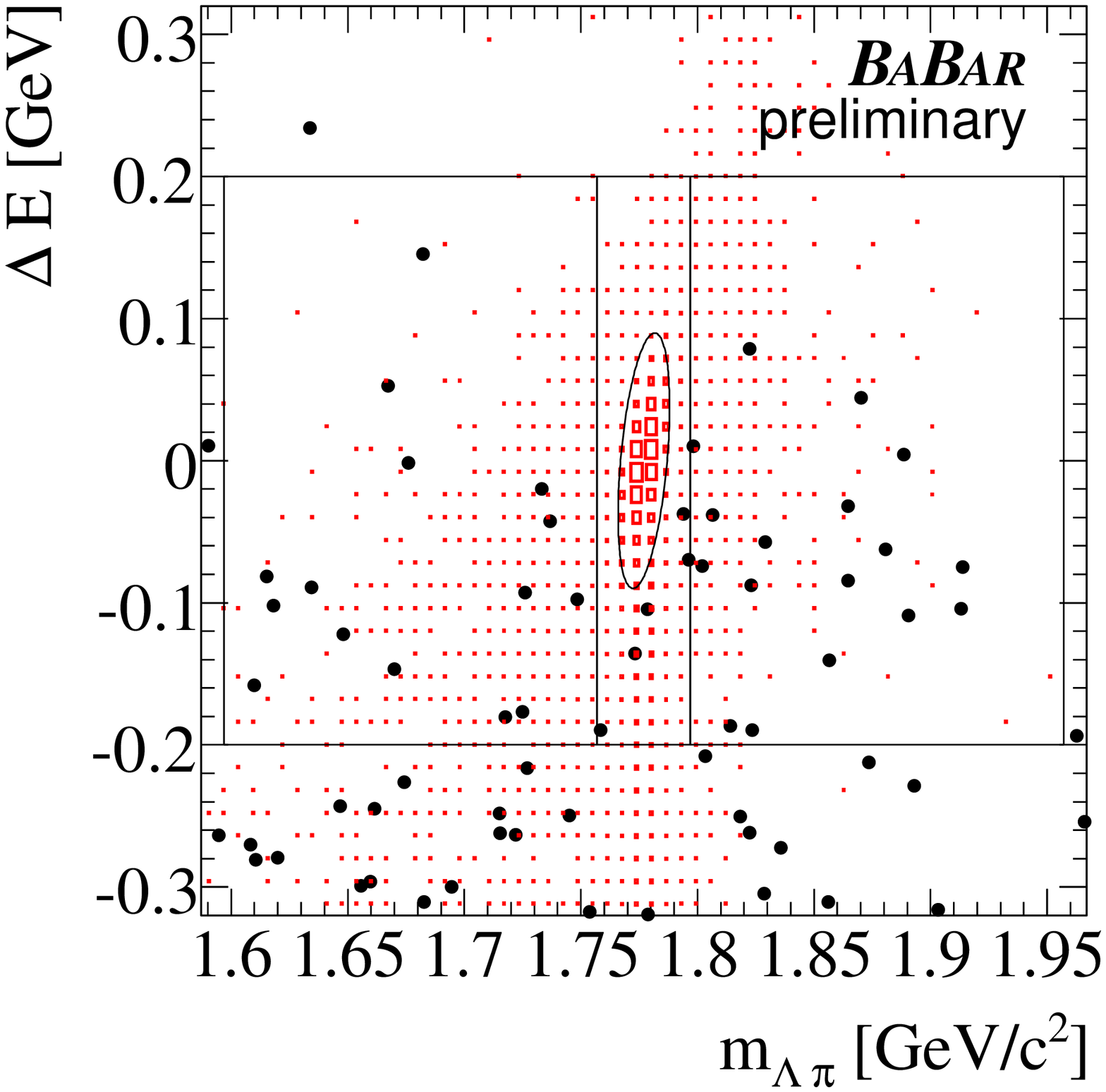,width=6cm}}
    \subfigure[$\tau^- \to \bar{\Lambda^0} K^-$]{\epsfig{file=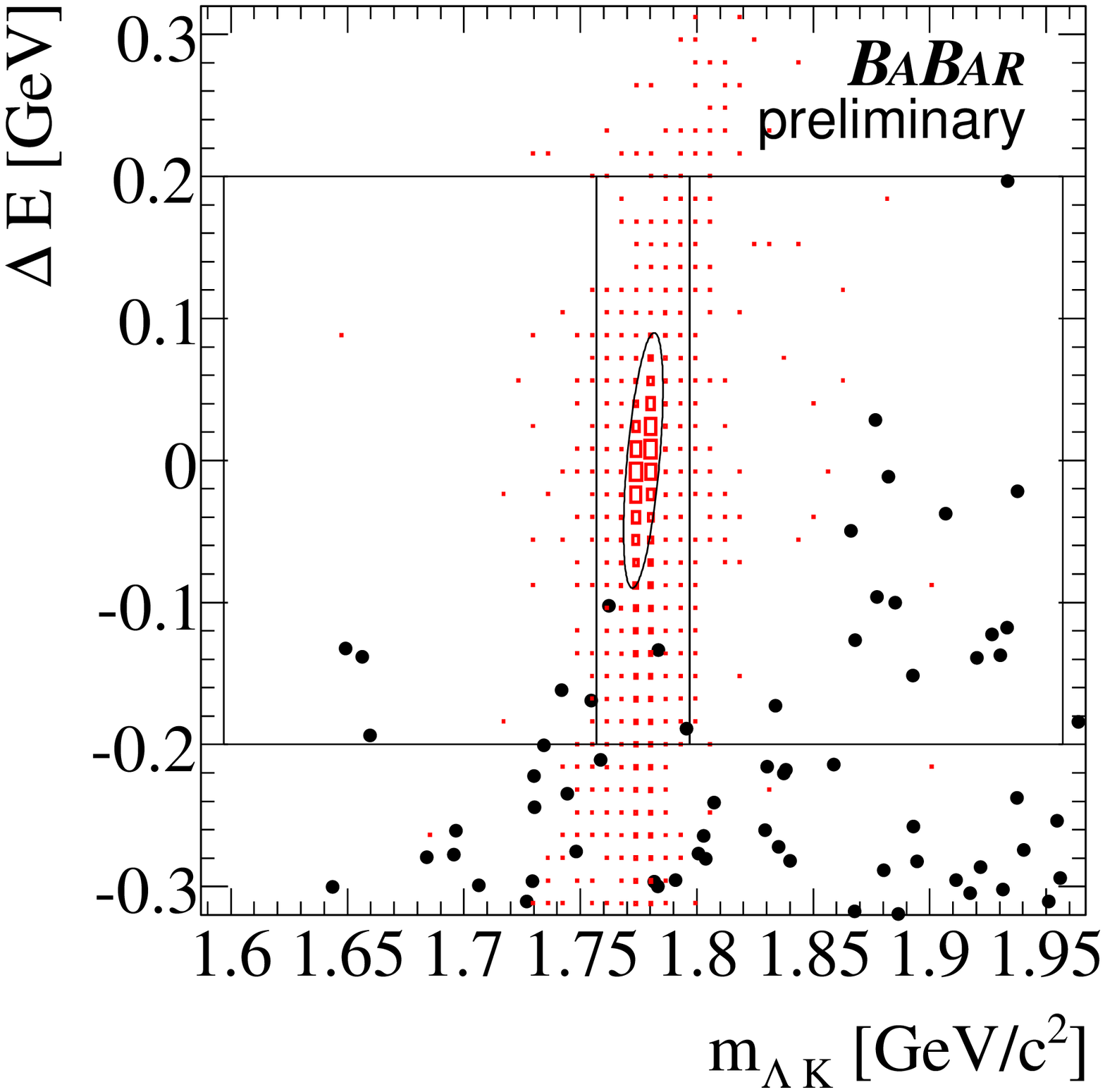,width=6cm}}
    \subfigure[$\tau^- \to \Lambda^0 K^-$]{\epsfig{file=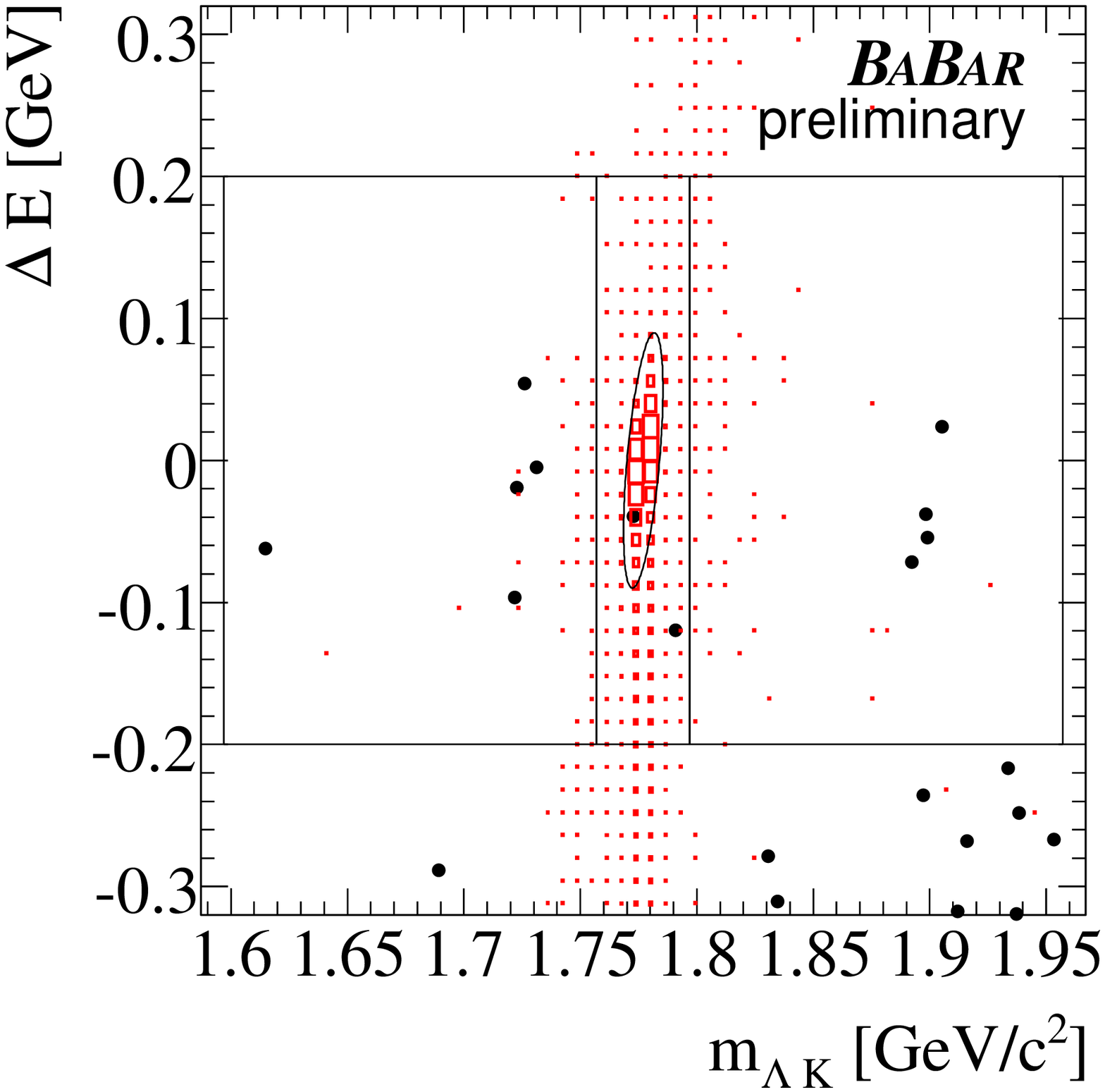,width=6cm}}
    \caption{$\Delta E_{\Lambda \pi(K)}$ versus $m_{\Lambda \pi(K)}$ data distributions for the $(B\!-\!L)$-conserving modes (left) and the $(B\!-\!L)$-violating modes (right).
    The top row shows the mode $\tau \to \Lambda \pi$; the mode $\tau \to \Lambda K$ is shown in the bottom row.
    The expected signal distribution (taken from Monte Carlo) is shown with red squares; data events are shown as dots. 
    The large rectangles in each plot are from left to right: left sideband, blinded region, and right sideband. 
    The elliptical signal region is also shown.}
    \label{figdEvsM2d}
  \end{center}
\end{figure}

The data distributions in the $\Delta E_{\Lambda \pi(K)}$ versus
$m_{\Lambda \pi(K)}$ plane after all selection requirements are shown
in Figure \ref{figdEvsM2d}.  No signal candidate events are observed
in the $\tau \to \Lambda \pi$ mode.  We observe one candidate event in
the $(B\!-\!L)$-violating mode $\tau^- \to \Lambda^0 K^-$.  We determine
upper limits on branching fractions at $90\%$ C.L.  using the method
described in Ref.~\cite{barlow}. This method considers uncertainties
both on the signal efficiency as well as the number of expected
background events in the signal region.  The number of expected
background events and number of observed events in the signal region,
the signal efficiency, and the upper limit that has been determined
are shown separately for the $(B\!-\!L)$-violating and $(B\!-\!L)$-conserving
cases in Table \ref{tblResults}. The upper limit on the branching
fraction is given by
\begin{equation}
{\cal B}_{U.L.}(\tau \to \Lambda \pi(K)) = \frac{\ell}{2 \sigma_{\tau\tau} {\cal L} {\cal B}(\Lambda \to p \pi) \varepsilon}\mbox{ ,}
\end{equation}
where $\ell$ is the $90\%$ C.L. upper limit for the signal yield,
$\sigma_{\tau\tau}=0.89$~nb is the assumed cross section for
production of $\tau$ pairs, ${\cal L}=237$~fb${}^{-1}$ is the total
luminosity of our dataset, ${\cal B}(\Lambda \to p \pi)=0.639$ is the
$\Lambda$ branching fraction taken from the RPP \cite{PDG}, and
$\varepsilon$ is the signal efficiency.

\begin{table}[!htb]
\begin{center}
\caption{The number of expected background events in the signal region, signal efficiency, number of observed events, $90\%$ C.L.
upper limit for the signal yield ($\ell$), and the upper limit branching fraction for each mode.\vspace{0.3cm}}
\begin{tabular}{lcccccc}
\hline \hline
mode & $(B\!-\!L)$ & expected   & efficiency &  observed & $\ell$ & upper limit on ${\cal B}$\\
     &         & background & \%         &  events   &        & @ 90\% C.L.\\
\hline\\[-0.25cm]
$\tau^- \to \bar{\Lambda^0} \pi^-$ & conserving & $0.42\pm 0.42$ &   $12.28$ & 0       & $1.97$ & $5.9 \times 10^{-8}$ \\
$\tau^- \to \Lambda^0 \pi^-$       & violating  & $0.56\pm 0.56$ &   $12.21$ & 0       & $1.90$ & $5.8 \times 10^{-8}$ \\
$\tau^- \to \bar{\Lambda^0} K^-$   & conserving & $0.26\pm 0.26$ &   $10.63$ & 0       & $2.08$ & $7.2 \times 10^{-8}$ \\
$\tau^- \to \Lambda^0 K^-$         & violating  & $0.12\pm 0.12$ &    $9.47$ & 1       & $3.78$ & $15 \times 10^{-8}$ \\
\hline \hline
\end{tabular}
\label{tblResults}
\end{center}
\end{table}

\section{SUMMARY}
\label{sec:Summary}

A search for the $(B\!-\!L)$-conserving modes $\tau^- \to \bar{\Lambda^0}
\pi^-$ and $\tau^- \to \bar{\Lambda^0} K^-$ as well as the
$(B\!-\!L)$-violating modes $\tau^- \to \Lambda^0 \pi^-$ and $\tau^- \to
\Lambda^0 K^-$ has been performed using $237$~fb${}^{-1}$ of $e^+e^-$
data. No signal is observed and we obtain preliminary upper limits on
the branching fractions at $90\%$ C.L. of ${\cal B}(\tau^- \to
\bar{\Lambda^0} \pi^-) < 5.9 \times 10^{-8}$, ${\cal B}(\tau^- \to
\Lambda^0 \pi^-) < 5.8 \times 10^{-8}$, ${\cal B}(\tau^- \to
\bar{\Lambda^0} K^-) < 7.2 \times 10^{-8}$, and ${\cal B}(\tau^- \to
\Lambda^0 K^-) < 15 \times 10^{-8}$. This analysis is the first
measurement of the mode $\tau \to \Lambda K$, and it improves over
earlier measurements of the mode $\tau \to \Lambda \pi$.

\section{ACKNOWLEDGMENTS}
\label{sec:Acknowledgments}

We would like to thank Yury Kamishkov for his insights and useful
discussions around baryon and lepton number violation in tau decays.
We are grateful for the 
extraordinary contributions of our \pep2\ colleagues in
achieving the excellent luminosity and machine conditions
that have made this work possible.
The success of this project also relies critically on the 
expertise and dedication of the computing organizations that 
support \babar.
The collaborating institutions wish to thank 
SLAC for its support and the kind hospitality extended to them. 
This work is supported by the
US Department of Energy
and National Science Foundation, the
Natural Sciences and Engineering Research Council (Canada),
Institute of High Energy Physics (China), the
Commissariat \`a l'Energie Atomique and
Institut National de Physique Nucl\'eaire et de Physique des Particules
(France), the
Bundesministerium f\"ur Bildung und Forschung and
Deutsche Forschungsgemeinschaft
(Germany), the
Istituto Nazionale di Fisica Nucleare (Italy),
the Foundation for Fundamental Research on Matter (The Netherlands),
the Research Council of Norway, the
Ministry of Science and Technology of the Russian Federation, and the
Particle Physics and Astronomy Research Council (United Kingdom). 
Individuals have received support from 
the Marie-Curie IEF program (European Union) and
the A. P. Sloan Foundation.

\end{document}